\documentclass[prl]{revtex4}
\usepackage{amsmath}
\usepackage{graphicx}
\usepackage{amsthm}
\newtheorem*{theorem}{Theorem}
\newtheorem{lemma}{Lemma}
\newtheorem{conjecture}{Conjecture}
\newcommand{\be}{\begin{equation}}
\newcommand{\ee}{\end{equation}}
\newcommand{\mO}{{\cal O}}
\begin{document}
\title{An Area Law for One Dimensional Quantum Systems}
\author{M.~B.~Hastings}
\affiliation{Center for Nonlinear Studies and Theoretical Division,
Los Alamos National Laboratory, Los Alamos, NM, 87545}
\begin{abstract}
We prove an area law for the entanglement entropy in gapped one dimensional
quantum systems.  The bound on the entropy grows surprisingly rapidly
with the correlation length; we discuss this in terms of properties of
quantum expanders and present a conjecture on completely positive
maps which may provide an alternate way of arriving at an area law.
We also show that, for gapped, local systems, the bound on
Von Neumann entropy implies a bound on R\'{e}nyi
entropy for sufficiently large $\alpha<1$
and implies the ability to approximate the ground state by a matrix product
state.
\end{abstract}
\maketitle

There are many reasons to believe that the entanglement
entropy of a quantum system with a gap
obeys an area law: that the entanglement entropy of a given
region scales as the boundary
area, rather than as the volume.
In one dimension,
conformal field theory calculations show that away from the critical
point
the entanglement entropy is bounded, diverging proportionally to the correlation
length as a critical point is approached\cite{cft}.
In higher dimensions, systems represented by matrix
product states\cite{mps,peps} obey an area law.

However, despite this, there is no general proof of an area law.
This is somewhat surprising, since it has been proven that correlation
functions in a gapped system decay exponentially\cite{loc}, and one might
guess that the decay of correlation functions implies that only degrees
of freedom near the boundary of the region may entangle with those outside.
However, the existence of data hiding states\cite{data} shows that one
can have states on bipartite systems with small correlations and large
entanglement.  Further, the existence of quantum expanders\cite{ent,exp} shows
that one may have matrix product states in one dimension that have
all correlation functions decaying exponentially in the distance between
the operators, but still have large entanglement.  This indicates
some of the difficulty in proving an area law.

At the same time, for a gapped system to violate an area law would require
some very strange properties.  For one thing\cite{solvgap}, the thermal
density matrix can be well approximated by a matrix product operator.
This implies that unless
a plausible assumption\cite{ent} on the density of low energy states
is violated, the ground
state can be well approximated by a matrix product state.

In this paper, we succeed in providing a proof of an area law for
one dimensional systems under
the assumption of a gap.  The result, however, bounds the entanglement
entropy by a quantity that grows exponentially in the correlation length.
This is much faster than the linear growth one might have expected.
We will comment later on why this bound might in fact be reasonably tight.
In the process of deriving this result, we will
derive bounds for gapped local systems which inter-relate three
quantities:
the Von Neumann entropy, the R\'{e}nyi entropy, and the error involved
in approximating
the ground state by a matrix product state.

We begin by defining the lattice and Hamiltonian.  We consider finite
range Hamiltonians for simplicity.  It is likely that the results can
be extended to exponentially decaying interactions, but for simplicity
we do not consider this here.  In fact, having decided to consider
only finite range interactions, we may group several sites into
a single site, and thus simplify to a problem with only
nearest neighbor interactions.

Specifically, we consider a finite volume one-dimensional
lattice, with sites labeled $i=1,2,...,N$, with a $D$-dimensional
Hilbert space on each site.
We consider finite-range
Hamiltonians of the form $H=\sum_{i=1}^N H_{i,i+1}$.
The finite range condition is that $H_{i,i+1}$ has support
on the set of sites $i$ and $i+1$.
We additionally impose a finite interaction strength condition bounding
the operator norm,
that the operator norm $\Vert H_{i,i+1} \Vert \leq J$ for some $J$.

The properties imply a Lieb-Robinson bound\cite{lr,hk,ns}: there exists
a velocity $v$ and length scale $\xi_C$
such that for any two operators $A,B$ with
support on sets $X,Y$ respectively,
\be
\label{lrb}
\Vert [A(t),B] \Vert \leq
c \times |X| \Vert A \Vert \Vert B \Vert
\exp[-\xi_C {\rm dist}(X,Y)]
\ee
for $|t|\leq l/v$
where the distance between sets $X,Y={\rm min}_{i\in X,j\in Y}(|i-j|)$,
where $c$ is a numeric constant of order unity,
and where $A(t)=\exp[i H t]
A \exp[-i H t]$.
The velocity $v$ will be of order $J$, while $\xi_C$ will be of order
unity.

We now introduce some notation.  We let $X_{j,k}$ denote the set
of sites $i$ with $j\leq i \leq k$.  We let $\Psi_0$ denote the
ground state of the Hamiltonian $H$ and we let $\rho^0_{1,N}=\Psi_0\rangle
\langle \Psi_0$ be the ground state density matrix.  We
let $\rho^0_{j,k}$ denote the reduced ground state density matrix
on the interval $X_{j,k}$.  That is, $\rho^0_{j,k}={\rm tr}_{i \not \in
X_{j,k}}(\rho^0_{1,N})$, where the partial trace is over sites not in $X_{j,k}$.
We define the entropy of any density matrix $\rho_{j,k}$ by
$S(\rho_{j,k})={\rm tr}_{i\in X_{j,k}}(\rho_{j,k}\ln(\rho_{j,k}))$.

\begin{theorem}
Consider a Hamiltonian satisfying the finite range and finite
interaction strength conditions above.  Suppose $H$ has a unique
ground state with a gap $\Delta E$ to the first excited state.
Then, for any $i$,
\be
\label{entbd}
S(\rho^0_{1,i})\leq S_{max}
\ee
where we define
\be
\label{smd}
S_{max}=c_0 \xi'\ln(\xi') \ln(D)2^{\xi' \ln(D)},
\ee
for some numerical constant $c_0$ of order unity,
and where we define
\begin{eqnarray}
\xi={\rm max}(2v/\Delta E,\xi_C), \\ \nonumber
\xi'=6\xi.
\end{eqnarray}
\end{theorem}
We do not consider the case of a degenerate ground state, but we expect
that the theorem can be strengthened to include this case also.

\section{Proof of Main Theorem}
To prove the theorem, we assume that it is false, so for some $i_0$,
we have
$S(\rho^0_{1,i})>S_{max}$.
Then, for any $k>i$ we have
$S(\rho^0_{1,k})>S_{max}-(k-i) \ln(D)$, and
therefore, for all $k$ with $i\leq k\leq i + l_{0}$
where
\be
\label{l0def}
l_{0}\equiv
S_{max}/3\ln(D),
\ee
we have 
\be
\label{sbnd}
S(\rho^0_{1,k})\geq 2 S_{max}/3 \equiv S_{cut}.
\ee
Then, define $S_l$, for each $l\leq l_0$, to be the maximum
entropy of an interval of length $l$ contained in $X_{i,i+l_0}$.
That is, $S_l$ is the maximum
of $S(\rho^0_{j+1,j+l})$
over $j$
such that $X_{j+1,j+l}\subseteq X_{i,i+l_0}$.
Clearly, 
\be
S_1\leq \ln(D).
\ee
Further, for any $l$, we have 
\be
\label{naibd}
S_{2l}\leq 2S_l.
\ee

However, it should be apparent that it is not possible for
Eq.~(\ref{naibd}) to be saturated.  If $S_{2l}=S_l$, then we have
for some $j$ that 
$S(\rho^0_{j-l+1,j+l})=S(\rho^0_{j-l+1,j})+S(\rho^0_{j+1,j+l})$.  
In this case,
the density matrix $\rho^0_{j-l+1,j+l}$ is equal to the product
of density matrices $\rho^0_{j-l+1,j}\otimes \rho^0_{j+1,j+l}$.
This implies that ${\rm tr}(H \rho^0_{1,j}\otimes \rho^0_{j+1,N})=
{\rm tr}(H \rho^0_{1,N})$ since $H$ is a sum of terms $H_{i,i+1}$ acting on
nearest neighbors.  However, given a
unique ground state, this implies that $\rho^0_{1,N}=\rho^0_{1,j}\otimes
\rho^0_{j+1,N}$,
contradicting the assumption of a non-vanishing entanglement
entropy
$S(\rho^0_{1,j})$.
The key idea of the proof will be to improve on Eq.~(\ref{naibd}) even
further, and use the fact that $S(\rho^0_{1,j})$ is not only non-vanishing,
but bounded below by $S_{cut}$
to show that the following stronger claim holds for all $l \leq l_0$:
\be
\label{claim}
S_{2l}\leq 2S_l-(1-2 C_1(\xi) \exp(-l/\xi')) l/\xi'+\ln(C_1(\xi))+C_2,
\ee
for some function $C_1(\xi)$ which
is bounded by a polynomial in $\xi$, and
some numerical constant $C_2$ of order unity.

Once Eq.~(\ref{claim}) is shown, we can iterate it.  
The dominant terms on the right-hand side of Eq.~(\ref{claim}) at large $l$
are $2S_l-l/\xi'$, so that one can guess from the behavior of
these dominant terms that we will find $S_l\approx l\ln(D)-\log_2(l/\xi_0)
l/\xi'$, up to subleading terms.  Indeed this is the case.
Set $\xi_0=\xi' 2 \ln(2C_1(\xi))$.  We use $S_l \leq l \ln(D)$ for $l=\xi_0$
as an initial bound and iterate to $l=2\xi_0,l=4\xi_0$, and so on.
Note that $2C_1(\xi)\exp[-\xi_0/\xi']\xi_0/\xi'\leq 1$,
and $\sum_{n=0}^{\infty}
2C_1(\xi)\exp[-2^n\xi_0/\xi']\xi_0/\xi'$ is bounded by
$1+e^{-2}+e^{-4}+...\leq 2$.
It follows that for $l>\xi_0$,
\be
S_l \leq \ln(D) l - l \lfloor{\log_2(l/\xi_0)}\rfloor/\xi'
+(2 + \ln(C_1(\xi)) + C_2) l/\xi_0.
\ee
However, since $S_l$ must be positive, a contradiction will
arise when $\lfloor\log_2(l/\xi_0)\rfloor\geq \ln(D)\xi'+(2+C_2)\xi'/\xi_0+1/2$
and so
\be
l_0 \leq \xi' \ln(2 C_1(\xi)) 2^{\lceil \xi' \ln(D)+(2+C_2)\xi'/\xi_0+1/2\rceil}.
\ee
Combined with Eq.~(\ref{l0def}) this gives the main result
for some constant $c_0$.

Thus, we must now show Eq.~(\ref{claim}).  To show this, it suffices
to show that for all $j,l$ with $X_{j+1,j+l}\subseteq X_{i,i+l_0}$
that,
\be
\label{relent}
S(\rho^0_{j-l+1,j+l})\leq
S(\rho^0_{j-l+1,j})+S(\rho^0_{j+1,j+l})-(1-2 C_1(\xi) \exp(-l/\xi'))
l/\xi'+\ln(C_1(\xi))+C_2.
\ee

At this point, we need two lemmas that rely on the Lieb-Robinson
bound and the existence of a gap.
\begin{lemma}
Let $H$ be a Hamiltonian that satisfies the finite range, finite
interaction strength, and gap conditions above.  Then for any $j$ and any
$l$
there exist Hermitian, positive definite,
operators $O_B(j,l),O_L(j,l),O_R(j,l)$ with the following properties.
First, $\Vert O_B(j,l) \Vert \leq 1$, 
$\Vert O_L(j,l) \Vert \leq 1$, and $\Vert O_R(j,l) \Vert \leq 1$.

Second,
\be
\label{prodapprox}
\Vert O_B(j,l) O_L(j,l) O_R(j,l)-P_0 \Vert\leq C_1(\xi) \exp[-l/\xi']\equiv \epsilon(l),
\ee
where $P_0=\Psi_0\rangle\langle\Psi_0$
is the projection operator onto the ground state and the function
$C_1(\xi)$ is bounded by a polynomial in $\xi$.
Finally, $O_L(j,l)$ is supported on $X_{1,j}$, $O_R(j,l)$ is supported
on $X_{j+1,N}$, and $O_B(j,l)$ is supported on $X_{j-l+1,j+l}$.

We prove this lemma in the appendix.
\end{lemma}

The next lemma provides a way of bounding the entanglement
entropy of a given region, given a certain assumption: that one
can approximate the ground state to a certain
accuracy in Hilbert space norm by a state with
given Schmidt rank.
In general, if one knows that a matrix product state provides a good
approximation to the ground state, this lemma
can be used to ``bootstrap" that result into a bound on the entanglement
entropy of the ground state.

\begin{lemma}
\label{ble}
Let
\be
\rho=\sum_{\gamma} P(\gamma) 
\sum_{\alpha=1}^k A(\alpha,\gamma) \Psi_L(\alpha,\gamma)\otimes
\Psi_R(\alpha,\gamma) \Bigr\rangle \Bigl\langle
\sum_{\beta=1}^k A(\beta,\gamma) \Psi_L(\beta,\gamma)\otimes
\Psi_R(\beta,\gamma)
\ee
be some density matrix with unit trace.
Here, $\Psi_L(\alpha,\gamma)$ are states on $X_{1,j}$ and
$\Psi_R(\alpha,\gamma)$ are states on $X_{j+1,N}$.
Then, we say that $\rho$ is a mixture of pure states with
Schmidt rank at most $k$.
Suppose that $\langle \Psi_0,\rho \Psi_0 \rangle=P>0$, where $\Psi_0$
is the ground state of a Hamiltonian that satisfies the finite range,
finite interaction length, and gap conditions above.
Then,
\be
\label{bootstrap}
S(\rho^0_{1,j}) \leq
\ln(k)+\xi'\ln(2C_1(\xi)^2/P)\ln(D)+F(\xi',D),
\ee
where
\be
F(\xi',D)=(\xi'+4)\ln(D)
+1+\ln(D^2-1)
+\ln(\xi'/2+1),
\ee
and a similar bound holds for R\'{e}nyi entropies with sufficiently
large $\alpha$ as discussed below (here, $\alpha$ refers to the order of the
R\'{e}nyi entropy, not to a particular state; the particular use
of $\alpha$ should be clear in context).

\begin{proof}
To prove this, 
note that for any
$m$ the positive definite operator
$\rho(m)=O_B(j,m) O_L(j,m) O_R(j,m) \rho O_R(j,m) O_L(j,m) O_B(j,m)$
has the following properties.
First, $\rho(m)$ is a mixture of pure states with Schmidt
rank at most $kD^{2m}$.
Second, 
from Eq.~(\ref{prodapprox}), ${\rm tr}(P_0 \rho(m))\geq
(\sqrt{P}-\epsilon(l))^2$ and ${\rm tr}((1-P_0)\rho(m))\leq \epsilon^2$
and thus
\be
\label{hg}
\frac{\langle \Psi_0,\rho(m),\Psi_0 \rangle}{{\rm tr}(\rho(m))} \geq
1-2C_1(\xi)^2\exp[-2m/\xi']/P.
\ee

Let the ground state $\Psi_0$ be equal to
$\sum_{\alpha=1}
A_0(\alpha) \Psi_{L,0}(\alpha)\otimes \Psi_{R,0}(\alpha)$,
where the states $\Psi_{L,0}(\alpha)$ and $\Psi_{R,0}(\alpha)$ 
are again states
on $X_{1,j}$ and $X_{j+1,N}$ respectively, with
$\langle \Psi_{L,0}(\alpha),\Psi_{L,0}(\beta)\rangle=
\langle \Psi_{R,0}(\alpha),\Psi_{R,0}(\beta)\rangle=\delta_{\alpha,\beta}$.
We order the different states such that
if $\alpha<\beta$ then $|A_0(\alpha)|\geq |A_0(\beta)|$ and
we normalize so that $\sum_{\alpha} |A_0(\alpha)|^2=1$.
It follows from Eq.~(\ref{hg}) that for all integer $m$
\be
\label{constraint}
\sum_{\alpha\geq kD^{2m}+1} |A_0(\alpha)|^2\leq
2C_1(\xi)^2\exp[-2m/\xi']/P.
\ee
Let $m'$ be the smallest integer $m$ such that $2 \exp[-2m/\xi']C_1(\xi)^2/P\leq 1$.
Thus, for $m>m'$
\be
\label{constraint2}
\sum_{\alpha\geq kD^{2m}+1} |A_0(\alpha)|^2\leq
\exp[-2(m-m')/\xi'].
\ee
We now maximize
the entropy
$S(\rho^0_{1,j+l})=-\sum_{\alpha=1} |A_0(\alpha)|^2 \ln(|A_0(\alpha)|^2)$
subject to the constraint (\ref{constraint2}).
The maximum occurs when
$\sum_{\alpha=1}^{kD^{2m'+2}} |A_0(\alpha)|^2=(1-\exp[-2/\xi'])$
and for $m>m'$,
$\sum_{\alpha=kD^{2m}+1}^{kD^{2m+2}} |A_0(\alpha)|^2=(1-\exp[-2/\xi'])
\exp[-2(m-m')/\xi']$, giving an entropy bounded by
\begin{eqnarray}
&&
\ln(k)+(2m'+2)\ln(D)
\sum_{n=1}^{\infty}
(2(n-1)\ln(D)+\ln(D^2-1)+2n/\xi')
\exp[-2n/\xi']
(1-\exp[-2/\xi'])
\\ \nonumber
&&-\ln(1-\exp[-2/\xi'])
\\ \nonumber
&=&
\ln(k)+(2m'+2+\frac{2\exp[-4/\xi']}{1-\exp[-2/\xi']})\ln(D)
+(2/\xi')\frac{\exp[-2/\xi']}{1-\exp[-2/\xi']}+\ln(D^2-1)\exp[-2/\xi']
-\ln(1-\exp[-2/\xi'])
\\ \nonumber
&\leq&
\ln(k)+(\xi'\ln(2C_1(\xi)^2/P)+\xi'+4)\ln(D)
+1+\ln(D^2-1)
+\ln(\xi'/2+1),
\end{eqnarray}
where we used the inequalities $2\exp[-4/\xi']/(1-\exp[-2/\xi'])\leq \xi'$,
$(2/\xi')\exp[-2/\xi']/(1-\exp[-2/\xi'])\leq 1$, and
$1/(1-\exp[-2/\xi'])\leq \xi'/2+1$
for $\xi'>0$
giving Eq.~(\ref{bootstrap}) as claimed.

We note that this proof can be extended to
R\'{e}nyi entropies $S_{\alpha}(\rho^0_{1,j}))$ defined by
$S_{\alpha}(\rho_{1,j})\equiv (1-\alpha)^{-1}\ln({\rm tr}(\rho_{1,j}^{\alpha}))$ for sufficiently large $\alpha$.
Maximizing the R\'{e}nyi entropy subject to the constraint (\ref{constraint})
gives $S_{\alpha}(\rho^0_{1,j})\leq
(1-\alpha)^{-1} \ln\{
\sum_{n=0}^{\infty} (kD^{2m'+2}D^{2n})^{1-\alpha} (1-\exp[-2/\xi'])^{\alpha}
\exp[-2n/\xi']^{\alpha}\}$.  The sum converges
so long as
\be
\exp[-2 \alpha/\xi'] (D^2)^{(1-\alpha)}<1,
\ee
in which case we have a
bound on the R\'{e}nyi
entropy which differs from the bound
bound (\ref{bootstrap}) on the von Neumann entropy only
in that the function $F(\xi',D)$ is replaced by an $\alpha$-dependent
function.
\end{proof}
\end{lemma}

We now return to proving the main theorem.
From Eq.~(\ref{prodapprox}) it follows that
$\langle \Psi_0,O_B(j,l) O_L(j,l) O_R(j,l) \Psi_0 \rangle
\geq 1-\epsilon(l)$ and hence
$\langle \Psi_0,O_B(j,l) \Psi_0 \rangle \times
\langle \Psi_0 O_L(j,l) O_R(j,l) \Psi_0 \rangle \geq (1-\epsilon(l))^2$.
Therefore, 
$\langle \Psi_0,O_B(j,l) \Psi_0 \rangle
\geq 1-2 \epsilon(l)$ and
$\langle \Psi_0,O_L(j,l) O_R(j,l) \Psi_0 \rangle
\geq 1-2 \epsilon(l)$.
Thus, 
\be
\label{expect1}
{\rm tr}(\rho^0_{j-l+1,j+l} O_B(j,l))\geq 1-2 \epsilon(l).
\ee

Set $P={\rm tr}(P_0 \rho^0_{1,j}\otimes \rho^0_{j+1,N})$.
From Eqs.~(\ref{sbnd},\ref{bootstrap}), we find that
$\xi'\ln(2C_1(\xi)^2/P)\ln(D)+F(\xi',D)
\geq S_{cut}$.
Therefore, $P\leq 2 C_1(\xi)^2 \exp[-(S_{cut}-F(\xi',D))/\ln(D) \xi']$.
Let 
\begin{eqnarray}
\label{expect2}
x&=&{\rm tr}(O_B(j,l) \rho^0_{1,j}\otimes \rho^0_{j+1,N})\\ \nonumber
&=&
{\rm tr}(O_B(j,l)\rho^0_{j-l+1,j}\otimes \rho^0_{j+1,j+l} ),
\end{eqnarray}
and
$y={\rm tr}(O_L(j,l) O_R(j,l) \rho^0_{1,j}\otimes \rho^0_{j+1,N})\geq
1-2\epsilon(l)$.
Then, $P \geq {\rm tr}(O_B(j,l) O_L(j,l) O_R(j,l) \rho^0_{1,j}
\otimes \rho^0_{j+1,N})-\epsilon(l)\geq
x y-\sqrt{x-x^2} \sqrt{y-y^2}
-\epsilon(l)$ as follows from a Cauchy-Schwarz inequality for the
expectation value of a product of operators.
Thus, $P \geq 
x (1-2\epsilon(l))-\sqrt{x-x^2} \sqrt{2 \epsilon(l)}
-\epsilon(l)\geq x(1-2 \epsilon(l))-\sqrt{x}\sqrt{2\epsilon(l)}
-\epsilon(l)$.
Thus,
\be
\label{xbd}
x\leq 
\{
2 C_1(\xi)^2 \exp[-(S_{cut}-F(\xi',D))/\ln(D) \xi']+
\sqrt{x}\sqrt{2\epsilon(l)}+2\epsilon(l)\}/(1-2\epsilon(l).
\ee

Thus, Eqs.~(\ref{expect1},\ref{expect2}) imply that the operator $O_B(j,l)$
has a large expectation value for the state
$\rho^0_{j-l+1,j+l}$ and a small expectation value for the
state
$\rho^0_{j-l+1,j}\otimes \rho^0_{j+1,j+l}$.
Then, the Lindblad-Uhlmann theorem\cite{lu} provides
a lower bound on the relative entropy
$S(\rho^0_{j-l+1,j+l}||\rho^0_{j-l+1,j}\otimes \rho^0_{j+1,j+l})$ giving
\begin{eqnarray}
\label{LU}
&& S(\rho^0_{j-l+1,j})+S(\rho^0_{j+1,j+l})-S(\rho^0_{j-l+1,j+l})\\ \nonumber
&\geq&
(1-2 \epsilon(l)) \ln((1-2\epsilon(l))/x)+2\epsilon(l)
\ln(2\epsilon(l)/(1-x)).
\end{eqnarray}

Eq.~(\ref{LU}) is the key step.
Everything that follows consists of picking the constant $C_2$
correctly.
We first assume that
$l_0\leq (S_{cut}-F(\xi',D))/\ln(D)-\xi' \ln(C_1(\xi))$.
If this assumption fails, then since $l_0=S_{cut}/2\ln(D)$,
we have $S_{cut}/2\ln(D)\leq F(\xi',D)/\ln(D)+\xi'\ln(C_1(\xi))$, and thus by
picking the constant $c_0$ large enough Eq.~(\ref{smd}) will
still hold.
Then, since $l\leq l_0$ it follows from this assumption that
$2 C_1(\xi)^2 \exp[-(S_{cut}-F(\xi',D))/\ln(D) \xi'] \leq 2 \epsilon(l)$.
Then from Eq.~(\ref{xbd}),
$x\leq (4\epsilon(l)+\sqrt{2\epsilon(l)x})/(1-2\epsilon(x))$.
Thus, as $l$ becomes large, we find that $x$ and
$\epsilon(l)$ approach zero exponentially.  Thus, for large enough $l$,
we have
$S(\rho^0_{j-l+1,j})+S(\rho^0_{j+1,j+l})-S(\rho^0_{j-l+1,j+l})\geq
(1-2\epsilon(l))\ln(1/x)$ plus some constant of order unity and
hence for all $l$ we have 
\be
S(\rho^0_{j-l+1,j})+S(\rho^0_{j+1,j+l})-S(\rho^0_{j-l+1,j+l})
\geq (1-2\epsilon(l)) \ln(1/\epsilon(l))-C_2,
\ee
where $C_2$ is a numeric constant of order unity.
This shows Eq.~(\ref{relent}) and completes the proof.

\section{Matrix Product States}
Our main result is the bound (\ref{entbd}).  We now use the
existence of a bound on entropy to construct
an approximation to the ground state by a matrix product state.
At first, this might
seem difficult, given that previous such constructions\cite{vc} relied
on the existence of a bound on the R\'{e}nyi entropy, and we have a bound
on the von Neumann entropy.  However ideas similar to those
used in the bootstrap lemma (\ref{ble})
will let us avoid these difficulties.

The construction applies to any system given the finite interaction
range, finite interaction strength, and gap conditions, and
given the existence of a bound $S(\rho^0_{1,j})\leq S_{max}$ for all $j$.
Consider a given $j$ and write the ground state as in lemma (\ref{ble}) by
$\Psi_0=\sum_{\alpha=1}
A_0(\alpha) \Psi_{L,0}(\alpha)\otimes \Psi_{R,0}(\alpha)$, with
$\Psi_{L,0}(\alpha),\Psi_{R,0}(\alpha)$ being states on $X_{1,j}$ and
$X_{j+1,N}$ respectively and the states ordered so that $|A_0(\alpha)|$ is
a non-increasing function of $\alpha$.

Suppose for some $k'$ we have
$\sum_{\alpha=k'+1}^\infty |A_0(\alpha)|^2>1/2$.  Then,
$|A_0(k'+1)|^2<1/2k$.  Thus, $S(\rho^0_{1,j})>(1/2) \ln(2k')$.
So, $k'>\exp(2S(\rho^0_{1,j}))/2$. Hence, for
\be
\label{k0Ent}
k_{0}=\exp(2S(\rho^0_{1,j}))/2
\ee
we have $\sum_{\alpha=1}^{k_0} |A_0(\alpha)|^2
\geq 1/2$.
At this point we could directly apply lemma (\ref{ble}) to bound the
R\'{e}nyi entropies, and thus get a matrix product form following
results in\cite{vc},
but we prefer to proceed more directly.
We apply Eq.~(\ref{constraint}) with $P=1/2$, getting
\be
\sum_{\alpha\geq k_0 D^{2m}+1} |A_0(\alpha)|^2\leq
4C_1(\xi)^2\exp[-2m/\xi'].
\ee

Therefore,
\be
\label{est}
\sum_{\alpha\geq k'} |A_0(\alpha)|^2 \leq 4C_1(\xi)^2
\exp(-\lfloor\log_D(k'/k_0)\rfloor/\xi')
\leq 4 C_1(\xi)^2 \exp(1/\xi') (k'/k_0)^{1/\xi'\ln(D)}.
\ee
This provides an estimate on how rapidly the Schmidt coefficients
decay and hence how accurately the ground state may be approximated
by a matrix product state.  In particular, for a chain of length
$N$, the error in approximating the ground state by a matrix product
state of bond dimension $k'$ scales as $N (k'/k_0)^{1/\xi'\ln(D)}$.

\section{Discussion and a Conjecture on Completely Positive Maps}
We now further explore the relationship between
a gap, exponentially decaying correlations, and an area law.  Consider
a one dimensional system with a gap.  We know that this state must have
exponentially decaying correlations.  However, this in itself does
not imply an area law.  For example, there exist matrix
product states
\be
\label{mps}
\Psi(s_1,s_2,...s_N)=\sum_{\alpha,\beta,...}
A_{\alpha,\beta}(s_1)
A_{\beta,\gamma}(s_2)
A_{\gamma,\delta}(s_3)...
\ee
with $s_i=1...D$
in which the associated completely positive map forms
a quantum expander so that a system with a low Hilbert space dimension
$D$ on each site and a short correlation length of order unity may have
a very large entropy\cite{ent}.  Consider, however, the following
model of a quantum expander modeled on that in \cite{ent}: we have a graph
with coordination number $D$
(which is a classical expander graph), where each node of the graph
corresponds to a value of the bond variable $\alpha$, each link from
one node to another gets labeled with one particular value of $s$, and
$A_{\alpha\beta}(s)$ is non-vanishing only if the link from $\alpha$
to $\beta$ is labeled with the given $s$.  The total
number of nodes in the graph is equal to the range of the bond variables,
and we will denote this number by $k$.  For a given value of $\alpha$,
the bond variable $\beta$ can assume $D$ different values, $\gamma$
can assume $D(D-1)$ different values, and so on, so that any correlations
between a bond variable $\alpha$ and another far away bond variable
which connects a distant pair of sites become small as required.
This behavior is shown in Fig.~1 for a system with $D=3$ and
an arbitrary possible choice of $\alpha,\beta,\gamma,\delta,\epsilon$.

\begin{figure}[tb]
\centerline{
\includegraphics[scale=0.7]{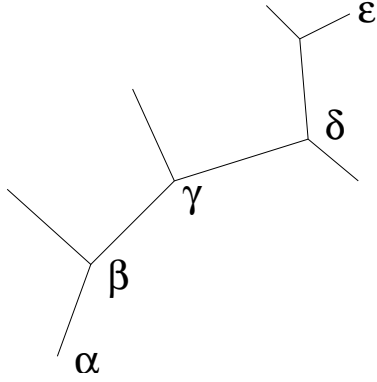}}
\caption{Illustration of an expander graph, with a particular choice of
$\alpha,\beta,...$ marked.  Note that each Greek index marks a site which
is one site away from the previous index.}
\end{figure}

However, as we have seen in this paper, if there is entropy across a
bond variable in a gapped, local system,
then there must be mutual information of order
$l/\xi'$ between the set
of sites within some distance $l$ to the left of the bond and the set of
sites within some distance $l$ to the right of the bond.  However,
in the given expander map, there is no such mutual information.
Consider a given set of sites $i=i_{min},i_{min}+1,...,i_{max}$,
with $i_{max}-i_{min}<<N$
and $D^{i_{max}-i_{min}}$ much smaller than the range $k$ of the
bond variable $\alpha$.  Trace out all
sites outside to construct a reduced density matrix on the given
sites: the result is proportional to the identity matrix and any of
the $D^{i_{max}-i_{min}+1}$ states is possible with equal probability.
Thus, this expander map is not the ground state of a Hamiltonian
with an energy gap of order unity and an interaction range of order
unity, even though it is the ground state of a gapped Hamiltonian with
an interaction range of order $\log_D(k)$.  In a sense, this expander
map is mixing {\it too} quickly to be the ground state of a gapped Hamiltonian.

There is, however, an interesting alternative way of seeing that
this expander map cannot be the ground state of a gapped local
Hamiltonian, which will lead to a conjecture we have on completely
positive maps.  We begin with a result on the decay of a certain kind
of correlation function.
\begin{lemma}
Let $H$ a Hamiltonian $H$ which satisfies
the finite range, finite interaction strength, and gap conditions.
Let $\Psi_0$ be the ground state of $H$.  Suppose
$\Psi_0=\sum_{\alpha=1} A_0(\alpha) \Psi_{L,0}(\alpha)\otimes
\Psi_{R,0}(\alpha)$, where $\Psi_{L,0}(\alpha)$ are orthonormal
states on $X_{1,j}$ and $\Psi_{R,0}(\alpha)$ are orthonormal states
on $X_{j+1,N}$.  Let
$B_L=\sum_{\alpha=1} O(\alpha) \Psi_{L,0}(\alpha) \rangle\langle \Psi_{L,0}
\otimes \openone_R$, where $\openone_R$ is the unit operator on $X_{j+1,N}$,
for some function $O(\alpha)$.
Similarly, let $B_R=
\openone_L \otimes
\sum_{\alpha=1} O(\alpha) \Psi_{R,0}(\alpha) \rangle\langle \Psi_{R,0}$.
Suppose $\Vert B_L \Vert \leq 1$, so that $|O(\alpha)|\leq 1$ for all $\alpha$.
Finally, let $A$ be an operator with support on $X_{1,j-l}\cup
X_{j+l+1,N}$ and with $\Vert A \Vert \leq 1$.
Then
\be
\label{fwdback}
\langle \Psi_0,A B_L \Psi_0 \rangle-\langle \Psi_0,A \Psi_0\rangle
\langle\Psi_0, B_L\Psi_0\rangle \leq
3\sqrt{2\epsilon(l)}+\epsilon(l),
\ee
where $\epsilon(l)$ is given as before by
$C_1(\xi) \exp[-l/\xi']$.
\begin{proof}
To prove this, define $O_B(j,l),O_L(j,l),O_R(j,l)$ as in Lemma 1.
Recall that $\langle \Psi_0,O_B(j,l) \Psi_0\rangle \geq 1-2\epsilon(l)$
and $\langle \Psi_0,O_R(j,l)\Psi_0 \rangle \geq 1-2\epsilon(l)$.
Hence, $|(O_B(j,l)-1)\Psi_0|\leq \sqrt{2\epsilon(l)}$ and
$|(O_R(j,l)-1)\Psi_0|\leq \sqrt{2\epsilon(l)}$.
We now use a series of triangle inequalities.
First,
\be
|\langle \Psi_0,A B_L \Psi_0 \rangle-
\langle \Psi_0,A B_L O_R(j,l) \Psi_0 \rangle| \leq \sqrt{2\epsilon(l)}.
\ee
Next,
\be
|\langle \Psi_0,A B_L O_R(j,l) \Psi_0 \rangle-
\langle \Psi_0,O_B(j,l) A B_L O_R(j,l) \Psi_0 \rangle|\leq
\sqrt{2\epsilon(l)}.
\ee
However, $[O_R(j,l),B_L]=0$ and $[O_B(j,l),A]=0$.
Also, $B_L \Psi_0=B_R \Psi_0$.
Thus, 
\be
\langle \Psi_0,O_B(j,l) A B_L O_R(j,l)\Psi_0 \rangle=
\langle \Psi_0,A O_B(j,l) O_R(j,l) B_R \Psi_0 \rangle.
\ee
Next, 
\be
|\langle \Psi_0,A O_B(j,l) O_R(j,l) B_R \Psi_0 \rangle-
\langle \Psi_0,A O_B(j,l) O_R(j,l) B_R O_L(j,l) \Psi_0 \rangle|\leq \sqrt{2\epsilon(l)}.
\ee
Using $[B_R,O_L(j,l)]=0$ we find that
\be
|\langle \Psi_0,A B_L \Psi_0 \rangle-
\langle \Psi_0,A O_B(j,l) O_R(j,l) O_L(j,l) B_R \Psi_0 \rangle|\leq 3\sqrt{2\epsilon(l)}.
\ee
Thus, 
$|\langle \Psi_0,A B_L \Psi_0 \rangle-\langle A P_0 B_L \Psi_0 \rangle|
\leq 3\sqrt{2\epsilon(l)}+\epsilon(l)$, completing the proof.
\end{proof}
\end{lemma}

Now, this lemma also implies that the matrix product state arising from
the given expander map cannot be the ground state of a Hamiltonian with
gap and Lieb-Robinson velocity of order unity.  In a slight abuse
of notation, 
let $\alpha$ be the bond variable which joins site $j-l$ to
$j-l+1$,
let $\gamma$ be the bond variable which joins site
$j$ to $j+1$, 
and let $\epsilon$ be the bond variable which joins site $j+l$ to
$j+l+1$.  This is shown in Fig.~1 for a system with $l=2$.

For any given value of the bond variable $\alpha$, the bond
variable $\gamma$
can have a wide range of possible values, of order $(D-1)^l$.
However, for given values of both $\alpha$ and $\epsilon$, the possible
range of values of $\gamma$ is much more restricted, as $\gamma$ will tend
to lie on the shortest path joining $\alpha$ to $\epsilon$
as shown in the figure.
In this manner, it is possible to construct
operators $A,B_L$ such that Eq.~(\ref{fwdback}) is violated for this state,
although we omit the detailed construction.

In fact, we have not been able to find any matrix product states which
obey Eq.~(\ref{fwdback}) while still having a large entanglement entropy.
We thus make the following conjecture on completely positive maps:
\begin{conjecture}
There exists a function $f(D_{eff})$ with the following property:
consider a matrix product state $\Psi_0$
as in Eq.~(\ref{mps}).  Suppose that this state satisfies Eq.~(\ref{fwdback})
for all $j,l,A,B_L$ for some given $\xi'$.  Then, the entropy $S(\rho^0_{1,j})$
is less than or equal to
$f(D^{\xi'})$.
\end{conjecture}
If this conjecture were shown, then it would give a different way to
prove an area law.

\section{Conclusion}
In conclusion we have given a proof of an area law, although the
bound is quite weak.
We note that the bootstrap lemma (\ref{ble}) gives some intuitive idea
as to why it is difficult to prove bounds on the entanglement entropy.
Given a good approximation to the ground state with a state of given
Schmidt rank, we have a bound on the ground state entropy.  Also, given
a bound on the ground state entropy, we can estimate how well we can
approximate the ground state with a state of given Schmidt rank as
in Eqs.~(\ref{k0Ent},\ref{est}).
However, this kind of argument leads to circular reasoning
and thus does not help provide an area law.
These arguments, do however, inter-relate the Von Neumann entropy,
the R\'{e}nyi entropy, and the error in approximating the ground
state by a matrix product state in gapped system, as in
Eqs.~(\ref{bootstrap},\ref{k0Ent},\ref{est}).

We finally consider the implications for numerical simulation of
one dimensional quantum systems.  If such a system has a gap, then
the ground state is close to a matrix product state.  However, finding
the best matrix product state may be hard problem\cite{npmp}.  Imagine,
though, that we consider a family of Hamiltonians which start from a Hamiltonian
with a known ground state and keep the gap open.  Then, by following a
quasi-adiabatic evolution\cite{qad} along this path and truncating the matrix
product state to keep the bond dimension bounded, we can well
approximate the ground state at the end of the evolution\cite{tjo}.

{\it Acknowledgements---} Portions of this work were completed at the
workshop on ``Lieb-Robinson Bounds and Applications" at the Erwin
Schr\"{o}dinger Institute.  I thank the other participants at the workshop
for useful discussions.
This work was supported by U. S. DOE Contract No. DE-AC52-06NA25396.

\section{Appendix: Approximating the Projection Operator}
In this appendix we show how to approximate the projection operator
by a product $O_B(j,l)O_L(j,l)O_R(j,l)$ as described in Lemma 1.
For simplicity, let us add a constant to $H$ so 
that the ground state energy is equal to zero.
We start by recalling the  result in \cite{qad}, that it is possible,
given a gapped Hamiltonian, to write the Hamiltonian as a sum of terms
such each term approximately annihilates the ground state.
We perform the derivation slightly differently to get a tighter
bound (\ref{tighter},\ref{tighter2}) which depends only on the
boundary of the terms.
Let 
\begin{eqnarray}
H_L=\sum_{i,i\leq j-l/3} H_{i,i+1}, \\ \nonumber
H_B=\sum_{i,j-l/3+1\leq i \leq j+l/3} H_{i,i+1}, \\ \nonumber
H_R=\sum_{i,i\geq j+1+l/3} H_{i,i+1},
\end{eqnarray}
so that $H=H_L+H_B+H_R$.
We then choose to add constants to $H_L,H_B,H_R$ so that $\langle
H_L\rangle=\langle H_B \rangle=\langle H_R\rangle=0$.
Define
\begin{eqnarray}
\tilde H_L^0=\frac{\Delta E}{\sqrt{2\pi q}} \int_{-\infty}^{\infty}
{\rm d}t\, H_L(t) \exp[-(t\Delta E)^2/2q], \\ \nonumber
\tilde H_B^0=\frac{\Delta E}{\sqrt{2\pi q}} \int_{-\infty}^{\infty}
{\rm d}t\, H_B(t) \exp[-(t\Delta E)^2/2q], \\ \nonumber
\tilde H_R^0=\frac{\Delta E}{\sqrt{2\pi q}} \int_{-\infty}^{\infty}
{\rm d}t\, H_R(t) \exp[-(t\Delta E)^2/2q],
\end{eqnarray}
where we pick $q=(l/3)\Delta E/(2v)$.
Note that $\tilde H_L^0+\tilde H_B^0+\tilde H_R^0=H$.
Let $b_L=[H,H_L]$.  Note that $\Vert b_L \Vert \leq J^2$.
Then
\begin{eqnarray}
\label{tighter}
|\tilde H_L^0 \Psi_0|& \leq & \Delta E^{-1} |H \tilde H_L^0 \Psi_0 |
=\Delta E^{-1} |[H,\tilde H_L^0]\Psi_0| \\ \nonumber
&=&\Delta E^{-1} |\frac{\Delta E}{\sqrt{2 \pi q}}
\int_{-\infty}^{\infty} {\rm d}t\, b_L(t) \Psi_0|
\\ \nonumber
&\leq & \Delta E^{-1} J^2 {\cal O}(\exp[-l/3\xi]),
\end{eqnarray}
where on the last inequality we used the assumption of a gap and
${\cal O}(...)$ is used to denote a bound up to a numeric constant
of order unity.
Similarly,
\begin{eqnarray}
\label{tighter2}
|\tilde H_B^0 \Psi_0| \leq {\cal O}(\Delta E^{-1} J^2 \exp[-l/3\xi]),
\\ \nonumber
|\tilde H_B^0 \Psi_0| \leq {\cal O}(\Delta E^{-1} J^2 \exp[-l/3\xi]).
\end{eqnarray}

Using the Lieb-Robinson bound, for the given value of
$q$ it is possible to approximate
$\tilde H_L^0,\tilde H_B^0,\tilde H_R^0$ by operators
$M_L,M_B,M_R$ respectively such that
$M_L-H_L$ is supported on $X_{j-2l/3,j}$,
$M_B-H_B$ is supported on $X_{j-2l/3,j+1+2l/3}$,
and $M_R-H_R$ is supported on $X_{j+1,j+1+2l/3}$ and
such that
$\Vert M_L-H_L \Vert\leq {\cal O}(\Delta E^{-1} J^2 \exp[-l/3\xi])$,
$\Vert M_B-H_B \Vert\leq {\cal O}(\Delta E^{-1} J^2 \exp[-l/3\xi])$,
$\Vert M_R-H_R \Vert\leq {\cal O}(\Delta E^{-1} J^2 \exp[-l/3\xi])$,

Thus, we have $|M_L \Psi_0| \leq {\cal O}(\Delta E^{-1} J^2 \exp[-l/3\xi])$,
and similarly for $M_B$ and $M_R$.
We now define $O_L(j,l)$ to project onto eigenvectors of $M_L$ with
eigenvalue less than or equal to $\Delta E^{-1} J^2 \exp[-l/6\xi]$,
and define
$O_R(j,l)$ to project onto eigenvectors of $M_R$ with
eigenvalue less than or equal to $\Delta E^{-1} J^2 \exp[-l/6\xi]$.
Thus, 
\begin{eqnarray}
\label{obd}
|(O_L(j,l)-1)\Psi_0|\leq {\cal O}(\exp[-l/6\xi]), \\ \nonumber
|(O_R(j,l)-1)\Psi_0|\leq {\cal O}(\exp[-l/6\xi]).
\end{eqnarray}
Clearly, $O_L(j,l)$ and $O_R(j,l)$ are supported as required by Lemma 1.

We now define an approximation to the projection operator
\be
P_{q}\equiv \frac{\Delta E}{\sqrt{2\pi q}} \int {\rm dt} \exp[i H t]
\exp[-(t\Delta E)^2/2q].
\ee
Using the spectral gap we have $\Vert P_q-P_0 \Vert \leq {\cal O}(\exp[-l/3\xi])$.
Thus, using $\Vert M_L+M_B+M_R-H \Vert \leq
{\cal O}(\Delta E^{-1} J^2 \exp[-l/3\xi])$, we have
\begin{eqnarray}
&&
\Bigl\Vert \frac{\Delta E}{\sqrt{2\pi q}}
\int {\rm dt} \exp[i (M_L+M_B+M_R) t]
\exp[-(t\Delta E)^2/2q]-P_0 \Bigr\Vert\\ \nonumber &=&
\Bigl\Vert \frac{\Delta E}{\sqrt{2\pi q}}
\int {\rm dt} \exp\Bigl[i \int_0^t {\rm d}t' \exp(i (M_L+M_R)t') M_B 
\exp(-i (M_L+M_R)t') \Bigr]\times \\ \nonumber &&
\exp[i (M_L+M_R)t]
\exp[-(t\Delta E)^2/2q]-P_0 \Bigr\Vert \\ \nonumber
&\leq &
{\cal O}(\Delta E^{-2} J^2\sqrt{q} \exp[-l/3\xi]),
\end{eqnarray}
where the exponential of the integral over $t'$ is $t'$-ordered.
Thus from Eq.~(\ref{obd}),
\begin{eqnarray}
\label{moved}
&& \Bigl\Vert \frac{\Delta E}{\sqrt{2\pi q}}
\int {\rm dt} \exp\Bigl[i \int_0^t {\rm d}t' \exp(i (M_L+M_R)t') M_B 
\exp(-i (M_L+M_R)t') \Bigr]
\times \\ \nonumber &&
\exp[i (M_L+M_R)t]
\exp[-(t\Delta E)^2/2q] O_L(j,l) O_R(j,l) -P_0 \Bigr\Vert \\ \nonumber
&\leq &
{\cal O}(\Delta E^{-2} J^2\sqrt{q} \exp[-l/3\xi]+\exp[-l/6\xi]).
\end{eqnarray}

However, 
$\Vert \exp[i (M_L+M_R)t] O_L(j,l) O_R(j,l)-O_L(j,l) O_R(j,l) \Vert \leq
2 |t| \Delta E^{-1} J^2 {\cal O}(\exp[-l/6\xi])$.
Combining this with Eq.~(\ref{moved}) we find that
\begin{eqnarray}
\label{last}
&& \Bigl\Vert \frac{\Delta E}{\sqrt{2\pi q}}
\int {\rm dt} \exp\Bigl[i \int_0^t {\rm d}t' \exp(i (M_L+M_R)t') M_B 
\exp(-i (M_L+M_R)t') \Bigr]\times \\ \nonumber
&& \exp[-(t\Delta E)^2/2q] O_L(j,l) O_R(j,l) -P_0 \Bigr\Vert \\ \nonumber
&\leq &
{\cal O}(\Delta E^{-2} J^2\sqrt{q} \exp[-l/6\xi]).
\end{eqnarray}
Consider the operator
\be
P_B\equiv \frac{\Delta E}{\sqrt{2\pi q}}
\int {\rm dt} \exp[i \int_0^t {\rm d}t' \exp(i (M_L+M_R)t') M_B 
\exp(-i (M_L+M_R)t') ]
\exp[-(t\Delta E)^2/2q].
\ee
Using a Lieb-Robinson bound for $M_L,M_R$ (and noting that the difference
$M_L+M_R-H_L-H_R$ has support on $X_{j-2l/3,j+1+2l/3}$), we can approximate
$P_B$ by an operator $O_B(j,l)$ with support on $X_{j-l+1,j+l}$ such that
$\Vert P_B-O_B(j,l) \Vert \leq {\cal O}(\Delta E^{-1} J \sqrt{q} \exp[-l/6\xi])$.
Thus,
\be
\Vert O_B(j,l) O_L(j,l) O_R(j,l) - P_0 \Vert \leq
{\cal O}(\Delta E^{-2} J^2\sqrt{l\Delta E/2v} \exp[-l/6\xi]=
{\cal O}(\sqrt{l/\xi} \xi^{2} \exp[-l/6\xi]).
\ee
This completes the result.

Note added:
Michael Levin has pointed out that this construction does not necessarily yield a Hermitian positive definite $O_B$ as claimed (the other properties do hold).
We will first explain why the rest of the proof goes through with only one minor change even with $O_B$ that is not positive definite, giving two distinct ways to do this; we finally explain how to ensure that $O_B$ is positive definite if desired.

While the application of Lindblad-Uhlmann needs a positive definite operator, we can simply consider the expectation
value of $O_B(j,l)^\dagger O_B(j,l)$ (which is manifestly positive definite) in $\rho_{j-l+1,j+l}$ and $\rho_{j-l+1,j}\otimes \rho_{j+1,j+l}$ rather than the expectation value of $O_B(j,l)$ when applying Lindblad-Uhlmann and the proof goes through as before as follows.
Since $\Vert O_B(j,l) O_L(j,l) O_R(j,l)-P_0 \Vert \leq \epsilon(l)$
then
$\Vert O_L(j,l) O_R(j,l) O_B(j,l)^\dagger O_B(j,l) O_L(j,l) O_R(j,l)-P_0 \Vert \leq 2\epsilon(l)+\epsilon(l)^2$.
Hence, the expectation value of $O_B(j,l)^\dagger O_B(j,l)$ is close to $1$ in the ground state (one can derive an identity similar to that between Eq. (21) and (22) that is) but close to
zero in the state
$\rho^0_{j-l+1,j} \otimes \rho^0_{j+1,j+l}$.
Thus, the proof goes through as before with only a replacement $\epsilon(l) \rightarrow 2\epsilon(l)+\epsilon(l)^2$.

In fact, even this is not necessary.  We can still consider the expectation value of $O_B(j,l)$ as done previously.
The real part of the expectation value is sensitive only to the Hermitian part of this operator, and indeed is close to $1$ in the ground state.
The fact that $O_B(j,l) O_L(j,l) O_R(j,l)$ is exponentially close to $P_0$ means that in the range of $O_L(j,l) O_R(j,l)$ the operator $O_B$ is exponentially close to $O_L(j,l) O_R(j,l) P_0 O_L(j,l) O_R(j,l)$ and so is exponentially close to Hermitian and positive definite within that subspace.
The ground state density matrix as well as the reduced density matrices $\rho_{1,j}$ and $\rho_{j_1,N}$ have only exponentially small projection outside this subspace.  Hence, Linbdlad-Uhlmann can still be applied.

However, it is of interest for other applications to show that we can in fact construct a Hermitian positive definite operator $O_B^+(j,l)$ for the same choice of $O_L(j,l),O_R(j,l)$ as in the lemma so that $O_B^+(j,l) O_L(j,l) O_R(j,l)$ is exponentially close to $P_0$; we add a superscript $+$ to emphasize that $O_B^+$ is positive semi-definite.
We will show that setting
$O_B^+(j,l)=O_B(j,l)^\dagger O_B(j,l)$ suffices.
We will use the fact that
$\Vert O_B(j,l) O_L(j,l) O_R(j,l)-P_0 \Vert \leq \epsilon(l)$
to show
that
$\Vert O_B(j,l)^\dagger O_B(j,l) O_L(j,l) O_R(j,l) - P_0 \Vert \leq \mO(\sqrt{\epsilon(l)})$.
This follows from the following lemma if we set $B=O_B(j,l), P=P_0$, and $Q=O_L(j,l) O_R(j,l)$.
The square-root epsilon dependence still gives the error an exponential decay with $l$.

\begin{lemma}
Let $B$ be an arbitrary (not necessarily Hermitian) operator with $\Vert B \Vert \leq 1$.  Let $P,Q$ be projectors.
Assume that
\be
\label{assumption}
\Vert B Q - P \Vert \leq \epsilon.
\ee
Then,
\be
\Vert B^\dagger B Q - P \Vert \leq \sqrt{1-(1-\epsilon)^2}+3 \epsilon+\epsilon^2=\mO(\sqrt{\epsilon}).
\ee
\begin{proof}
From Eq.~(\ref{assumption}), we have
\begin{eqnarray}
\label{ex0}
\Vert Q B^\dagger B Q - P \Vert &=& \Vert \Bigl(P + (Q B^\dagger - P) \Bigr) \Bigl(P+(BQ-P)\Bigr) -P\Vert
\\ \nonumber
&\leq & 2 \epsilon+\epsilon^2.
\end{eqnarray}

Also,
\begin{eqnarray}
\label{ex1}
\Vert (1-Q) B^\dagger B Q \Vert & = & \Vert (1-Q) B^\dagger P + (1-Q) B^\dagger (BQ-P) \Vert
\\ \nonumber
&\leq & \Vert (1-Q) B^\dagger P \Vert + \epsilon.
\end{eqnarray}
Since $\Vert B^\dagger \Vert \leq 1$, we have
\be
|(1-Q) B^\dagger P \psi|^2 + |Q B^\dagger P \psi|^2 \leq 1
\ee
for any $\psi$ with $|\psi|=1$.
Restricting without loss of generality to $\psi$ in the range of $P$ we have
$|Q B^\dagger P \psi|=|Q B^\dagger \psi|\geq |P \psi| - |(QB^\dagger - P)\psi| \geq 1-\epsilon$.
So,
for any $\psi$ (not necessarily in the range of $P$) with $|\psi|=1$ we have
\be
\label{ex2}
|(1-Q)B^\dagger P \psi| \leq \sqrt{1-(1-\epsilon)^2}.
\ee

So, from Eqs.~(\ref{ex1},\ref{ex2}) we have
\be
\label{ex3}
\Vert (1-Q) B^\dagger B Q \Vert \leq \sqrt{1-(1-\epsilon)^2}+\epsilon.
\ee
So, by Eqs.~(\ref{ex0},\ref{ex3}) we have
\be
\Vert B^\dagger B Q - P \Vert \leq \sqrt{1-(1-\epsilon)^2}+3 \epsilon+\epsilon^2.
\ee
\end{proof}
\end{lemma}
Remark: under these assumptions, the $\mO(\sqrt{\epsilon})$ dependence of the error is the best we can get.  Consider
\be
P=Q=\begin{pmatrix} 1 & 0 \\0 & 0 \end{pmatrix}
\ee
and
\be
B=\begin{pmatrix} 1-\epsilon & \sqrt{1-(1-\epsilon)^2} \\ 0 & 0 \end{pmatrix}.
\ee

\end{document}